\documentclass[apj]{emulateapj}

\usepackage{color}
\usepackage{graphicx}
\usepackage{graphicx,epstopdf}
\usepackage{enumerate}
\usepackage[bookmarks=true, pdfstartview={FitH},hyperfootnotes=false]{hyperref}

\newcommand{\elem}[1]{{\scriptsize~{#1}}}

\newcommand{\elesm}[2]{\textrm{{\scriptsize {#1}}{\tiny{#2}}}}

\shorttitle{Studying the ISM with MBM36 observations}
\shortauthors{Ursino, E., Galeazzi, M., Liu, W.}

\begin{document}
\title{Studying the Interstellar Medium and the inner region of NPS/Loop 1 with shadow observations toward MBM36}


\author{Ursino, E., Galeazzi, M.\altaffilmark{1}, and Liu W.}
\affil{Physics Department, University of Miami, Coral Gables, FL 33155}
\altaffiltext{1}{corresponding author, galeazzi@physics.miami.edu}


\begin{abstract}

We analyzed data from a shadow observation of the high density molecular cloud MBM36 ($l\sim4^\circ, b\sim35^\circ$) with \emph{Suzaku}. MBM36 is located in a region that emits relatively weakly in the 3/4~keV band, compared to the surrounding NPS/Loop 1 structure and the Galactic Bulge. The contrast between a high and low density targets in the MBM36 area allows one to separate the local and distant contributors to the Soft Diffuse X-ray Background, providing a much better characterization of the individual components compared to single pointing observations. We identify two non-local thermal components, one at $kT\approx0.12$~keV and one at $kT\approx0.29$~keV. The colder component matches well with models of emission from the higher latitude region of the Galactic Bulge. The emission of the warmer component is in agreement with models predicting that the NPS is due to a hypershell from the center of the Milky Way. Geometrical and pressure calculations rule out a nearby bubble as responsible for the emission associate with the NPS. Any Galactic Halo/CircumGalactic Halo emission, if present, is outshined by the other components. We also report an excess emission around $0.9$~keV, likely due to an overabundance of Ne\elem{IX}.

\end{abstract}



\section{Introduction}
\label{introduction}
The all-sky $ROSAT 3/4$~keV map \citep{Snowden95, Snowden97} shows several extended features on top of a rather uniform background. The largest of these emitting regions are towards the Galactic Bulge (from $\sim-20^\circ$ to $\sim20^\circ$) and towards the North Polar Spur (NPS). The NPS is the brightest arc of the radio Loop I and extends from $(l,b)\sim(30^\circ,8^\circ)$ to $(l,b)\sim(30^\circ,75^\circ)$.

When it was first identified \citep{Berkhuijsen71}, Loop I was associated with the shell of a SuperBubble (SB) with center $(l,b)\approx(329^\circ,+17^\circ.5)$ and radius $\approx 4^\circ$. The emission of the shell is expected to be produced by gas and dust expanding driven by supernova or stellar wind from the Sco-Cen OB association (centered at $\sim170$~pc from the Sun). The NPS, in particular, should be the brightest region of an old ($\sim 10^6$~years) Supernova Remnant (SNR). Combining radio data with the X-ray data from \emph{ROSAT} and with $N_H$ maps, \citet{Egger95} identify the NPS as the region of interaction between the Local Hot Bubble (LHB) and the shock waves from $\sim2\times10^5$ year-old SuperNovae in the Sco-Cen association. Similarly, \citet{Wolleben07} models the Loop I emission as due to the interaction between two bubbles less than 100~pc away. All these models are characterized by a local source for the radio/X-ray emission, where the Loop I/NPS is at a distance of about 100~pc, and possibly there is a counterpart in the Southern Galactic Hemisphere.

According to a very different model \citep{Sofue00}, instead, the Loop I emission comes from biconical hypershells. In this case the shell is originated by some starburst or explosive event occurred about 15~Myr ago in the region of the Galactic Center. Since the Galactic Center is $\sim8$~kpc away, the hypershell should have a size of several kpc. In recent years, indeed, observations at different wavelengths accumulated evidence in favor of an hypershell scenario. The \emph{Fermi Bubbles} \citep{Su10,Dobler10,Ackermann14} are large scale (order of some kpc) $\gamma$-ray features broadly overlapping Loop I. This region is also characterized by the WMAP haze in K-band microwaves \citep{Finkbeiner04, Dobler08} and by polarized synchrotron radiation at 2.3~GHz \citep{Carretti13}, in both case showing biconical structures that can arise from an active Galactic Center. The biconical model is supported also by UV observations \citep{Fox15}. An additional argument in support of the hypershell model comes form extragalactic observations. Several galaxies, in fact, show outflows  \cite{Duric83, Nakai87, Bland88, Shopbell98, Pietsch00, Cecil01, Cecil02, Ohyama02, Veilleux02} similar to the one predicted by \citet{Sofue00} for the Milky Way.

There are now a few X-ray observations performed in the NPS region, in order to understand its nature. Using three \emph{XMM-Newton} observations performed along the length of the NPS, \citet{Willingale03} claimed that the emission is of local origin. According to their model, the NPS is a region at the edge of a spherical cavity located at a distance $\sim210$~pc and of radius $\sim140$~pc, filled with a $kT=0.3$~keV plasma only partially absorbed by a wall between the LHB and the cavity. A \emph{Suzaku} observation, instead, found that the NPS is particularly rich in Nitrogen \citep{Miller08}. This effect is hard to explain with a local origin of the NPS emission related to the Sco-Cen association. The N excess is possibly due, instead, either to local enrichment occurred at a previous stage or to some non-local source associated with the hypershell model. \citet{Kataoka13} performed a large set of \emph{Suzaku} observations crossing the northern and southern edges of the \emph{Fermi Bubbles}, therefore probing several fields within the NPS. They found strong X-ray absorption, hardly explained with a local origin, but rather suggesting that the NPS emission comes from gas associated with strong past activity of the Galactic Center. An improved version of this work \citep{Kataoka15}, including \emph{Suzaku} and \emph{Swift} observations, has been published while finalizing our paper and brings similar conclusions. Similarly, \citet{Tahara15} analyzed several \emph{Suzaku} observations performed close to the edges of the \emph{Fermi Bubbles} and found an ubiquitous thermal component at $kT\approx0.3$~keV and a possible additional component at $kT\approx0.7$~keV. 

\begin{table*}
\begin{center}
\caption{Most important properties of the On-cloud and Off-cloud observations.
\label{tab_targets}}
\begin{tabular}{|l|c|c|c|c|c|c|c|c|}
\hline
Target & ID & RA & DEC & $l$ & $b$ &Start Date & Livetime & $N_H$ \\
       &    & (h:m:s) & (d:m:s)  &deg &deg & & (s) &($10^{20}$~cm$^{-2}$) \\ 
\hline
MBM36-ON & 508079020 &15 53 32.47 &-04 53 33.4& 3.9 &35.6 &2013-08-27T16:46:10& 61823&24.7\\
MBM36-OFF & 508074010 &15 55 43.30 &	-01 47 16.1 & 7.4&37.1 &  2013-08-29T21:23:57& 58685&6.8 \\
\hline
\end{tabular}
\tablecomments{The table columns report the target name, \emph{Suzaku} observation ID, nominal satellite pointing direction, observation start time, observation livetime, and the Hydrgone column density estimated using \emph{IRAS100} data.}
\end{center}
\end{table*}

Recently, \citet{Puspitarini14} compared the X-ray emission towards the Loop I/NPS region with 3D maps of the distribution of the Interstellar Medium (ISM) up to a distance of $\sim200$~pc. The ISM distribution does not show evidence neither of a nearby large SB interacting with the LHB, nor of strong absorbers next to the galactic plane to justify the interruption of the NPS below $b=8^\circ$. If the NPS is due to a bright region of a large shell, the associated cavity must be located beyond 200~pc from the Sun. There is indeed a cavity at $\sim150$~pc that, if filled with a million degree temperature plasma, could be responsible of the X-ray emission in the NPS, but it is not large enough to be associated with the whole Loop I structure. 

Investigating the NPS in X-rays is not an easy task, since there are several competing sources along the line of sight, all with comparable intensity. Locally we have the Solar Wind Charge Exchange (SWCX, generated within $\approx10$~A.U.) and the LHB (within 100~pc). Among the non-local sources we account for the Galactic Halo (GH), the Galactic Bulge (GB), and unresolved extragalactic sources. 

The GB, in particular, is the brightest feature in the central region (within $30^\circ$ of the Galactic Center) of the 3/4 keV \emph{ROSAT} maps. Above and below the Galactic Center, the GB is much brighter than the GH also at high latitude \citep{Snowden97} and is likely to be a major component to the X-ray emission in the MBM36 region.

Shadow observations are performed by comparing two nearby regions, one characterized by large absorption column (i.e. an intervening molecular cloud) and the other one with much less absorption. The contrast between the two regions allows the observer to separate between the emission coming from sources located in front and behind the strong absorber. In particular, using a molecular cloud located right outside the LHB, it is possible to disentangle the local from the non-local components. This technique, in the X-rays, has been successfully used first with \emph{ROSAT} \citep{Kuntz00}, then with \emph{Chandra, XMM-Newton} and \emph{Suzaku} (see \citealt{Gupta09} for a review) to investigate the nature of LHB, SWCX, and GH.

MBM36 (also known as LDN134 or MLB40) is a molecular cloud conveniently situated right outside the LHB, at $\approx 140$~pc from the Sun \citep{Juvela12}. It is at a latitude ($b\sim35^\circ$) relatively larger than the Galactic Bulge so that the surrounding X-ray background is not dominated by it. Moreover, it is located $\approx 20^\circ$ away from the NPS, inside the perimeter of Loop I: if the NPS is associated with a shell (either from SB or hypershell), MBM36 would allow us to probe the region covered by the associated ``bubble''. The shadow technique, in fact, allows for a much better characterization of the emission from local and non-local components compared to previous investigations. Moreover, while not probing directly the NPS, it allows us to verify if there is indeed some bubble (on local or large scale) that can also generate the NPS arc-like shape.

In this paper we describe the investigation of the ISM and GH using MBM36 to perform shadow observations. In particular we attempt to characterize the nature of the region enveloped by the Loop I and verify if it is local, non-local, or if the gas emitting the X-ray NPS feature is in fact isolated, separated from the Loop I structure. In section~\ref{Observation} we describe the targets of the observation and the data reduction. In section~\ref{analysis} we describe the models adopted and the spectral analysis. We discuss the results in  section~\ref{discussion} and conclude in section~\ref{conclusion} with a summary of our results.

Unless differently specified, for the rest of the paper we will generally dub with SB a SuperBubble model, without reference to a local (as described by \citealt{Berkhuijsen71} and \citealt{Egger95}) or a large scale nature (like the hypershell model of \citealt{Sofue00}).

\section{Observation and Data Reduction}
\label{Observation}

\begin{figure}
\plotone{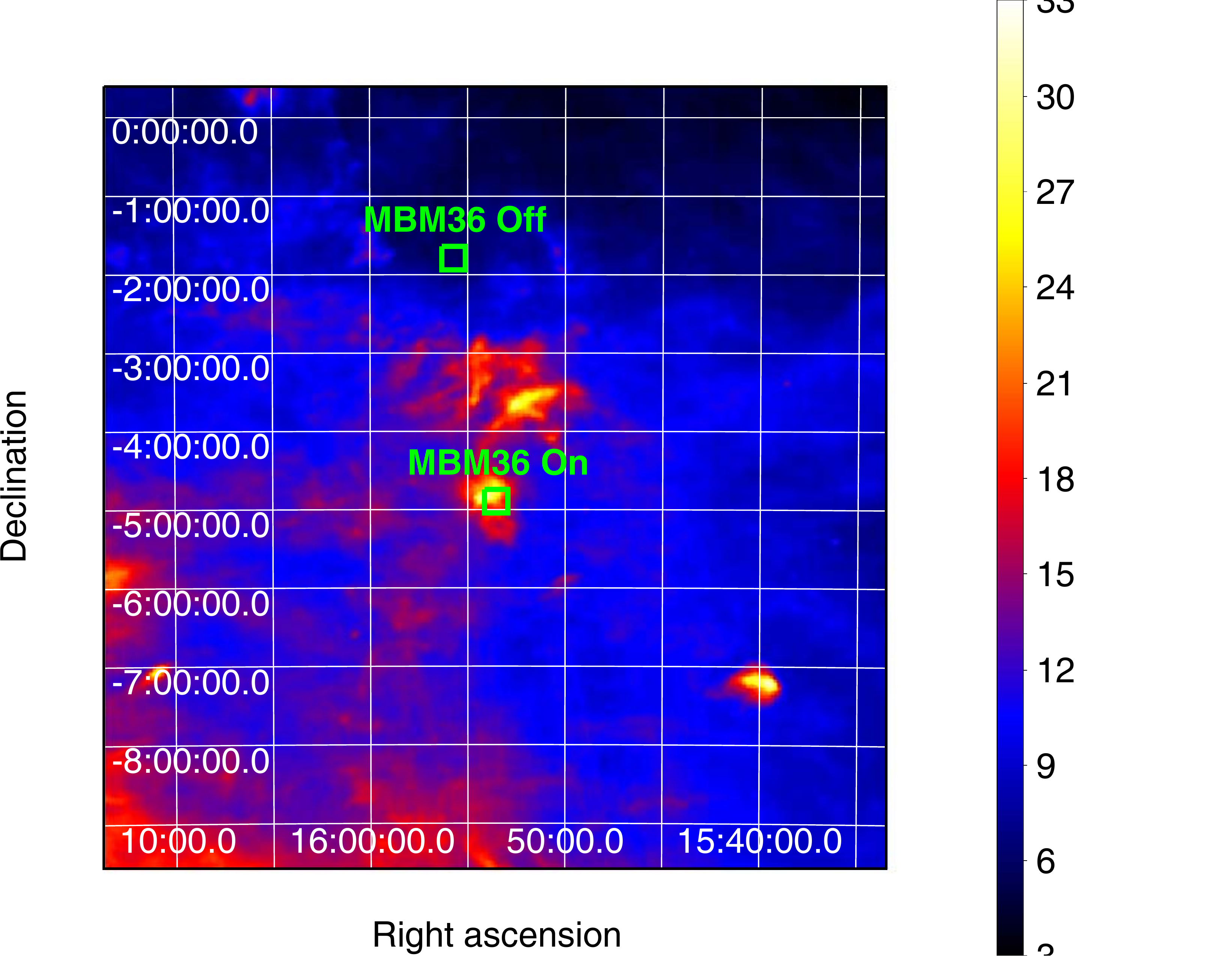}
\caption{\emph{IRAS100} map of the region around MBM36. The green boxes outline the On- and Off-cloud targets. Units are in MJy~sr$^{-1}$. 
\label{map_targets}}
\end{figure}

MBM36 is one of the targets of a \emph{Suzaku} SWCX key project, and \emph{Suzaku} performed several observations of the highest column density region of MBM36 in the past years. The key project required that the aimpoint was centered on the high density region of the cloud, since the \emph{Suzaku} field of view is too small to cover the whole cloud. In order to make full use of the shadow technique, we obtained a dedicated observation a few degrees away from MBM36, where the absorption column density is particularly low. We also required that the observations toward the on-cloud and the off-cloud regions were performed consecutively. The time constraint is essential to minimize the variation of the SWCX, usually on the order of a few days. For this analysis, we only used the pointing toward the dense region of MBM36 which is close, in time, to the low density pointing. For the remainder of the paper we refer to the two pointings as ``On-cloud'' and ``Off-cloud'' (or simply ``On''  and ``Off'').

Figure~\ref{map_targets} shows the position of the two targets overlapped to an \emph{IRAS100} map. The main parameters at the aimpoint of the observed fields are reported in table~\ref{tab_targets}. The two targets are separated by $\sim3^\circ$, this is a compromise between finding a region with the lowest possible absorption (and therefore maximizing the contrast with the On-cloud target) and staying as close as possible to the On-cloud (to minimize the variations of the medium probed). 

Surveys designed to measure the Galactic neutral hydrogen column density $N_H$ \citep{Dickey90, Kalberla05} lack the angular resolution necessary to resolve and characterize the absorption column towards the targets. \citet{Juvela12} performed a high resolution mapping of the molecular hydrogen column density in MBM36, unfortunately there is no ubiquitus way to convert the values of $N_{H_2}$ to $N_{H}$. In order to compute $N_H$, instead, we used the DIRBE-corrected \emph{IRAS100} maps. \emph{IRAS} maps have excellent angular resolution and allow us to clearly resolve the individual regions within the cloud (see figure~\ref{map_targets}). We converted the DIRBE-corrected \emph{IRAS100} values into column densities using the method described in \citet{Snowden00} for the northern galactic hemisphere, with the caveat that it has been tested only at relatively low column density (i.e. not in the density range of molecular clouds). Given the uncertainties in the on-cloud column density, we left it as a free parameter in the fit, although we also tested models fixed at the value reported in table~\ref{tab_targets}.

To reduce the dataset we used the dedicated \emph{Suzaku} software. First we reprocessed the raw data (\emph{aepipeline}) with updated calibration files, than we merged the data in $5\times5$ and $3\times3$ mode (\emph{xis5x5to3x3} and \emph{ftmerge}) in order to increase the sample. We filtered the data using the \emph{xselect} package. We first generated a clean dataset and than a map in the [$0.4-2.0$]~keV (channels $110-547$). We used the map to identify point sources (\emph{CIAO wavdetect} tool, with a reference flat map to reproduce 120'' PSF). We extracted the spectrum from the field of view (after removing the point sources and screening for the cut-off rigidity of the Earth's magnetic field with $COR2>8$~GeV) and generated the response, exposure, ancillary, and non X-ray background with the \emph{ftools xisrmfgen, xisexpmapgen, xissimarfgen}, and \emph{xisnxbgen}. Notice that we generated the response function assuming a uniform circular field of 20' of radius. Since the XIS0 and XIS3 (forward illuminated chips) have too few counts, we decided to analyze only data from the XIS1 chip, in order to reduce noise. We rescaled the non-X-ray background to the high energy tail of the spectrum ($11-14$~keV).

\emph{Suzaku} observations, in particular after 2011, have been increasingly affected by the O\elem{I} contamination at 0.525~keV \citep{Sekiya14}, caused by fluorescence of solar X-rays with neutral Oxygen in the Earth's atmosphere. Due to its proximity to the O\elem{VII} K-$\alpha$ triplet (between $0.560$~eV and $0.575$~eV) and the relatively poor energy resolution of the XIS detectors ($\sim0.1$~keV), O\elem{I} can strongly affect the spectral analysis. \citet{Sekiya14} showed that this effect can be minimized removing events taken during time intervals when the elevation angle from the bright Earth limb (the \emph{DYE\_ELV} parameter) is larger than $60^\circ$. The price to pay for this choice, however, is a much lower number of events and, for low count rate investigations, the statistics is greatly affected. Therefore, instead of using a stronger \emph{DYE\_ELV} filtering, we decided to include the O\elem{I} line in our models (see section~\ref{analysis}).

\section{Analysis}
\label{analysis}

For the analysis of the observed data, we used the XSPEC\footnote{http://heasarc.gsfc.nasa.gov/xanadu/xspec/} package, version 12.8.11 \citep{Arnaud96}, adopting the metal abundance model by \citet{Anders89}.

\begin{figure}
\plotone{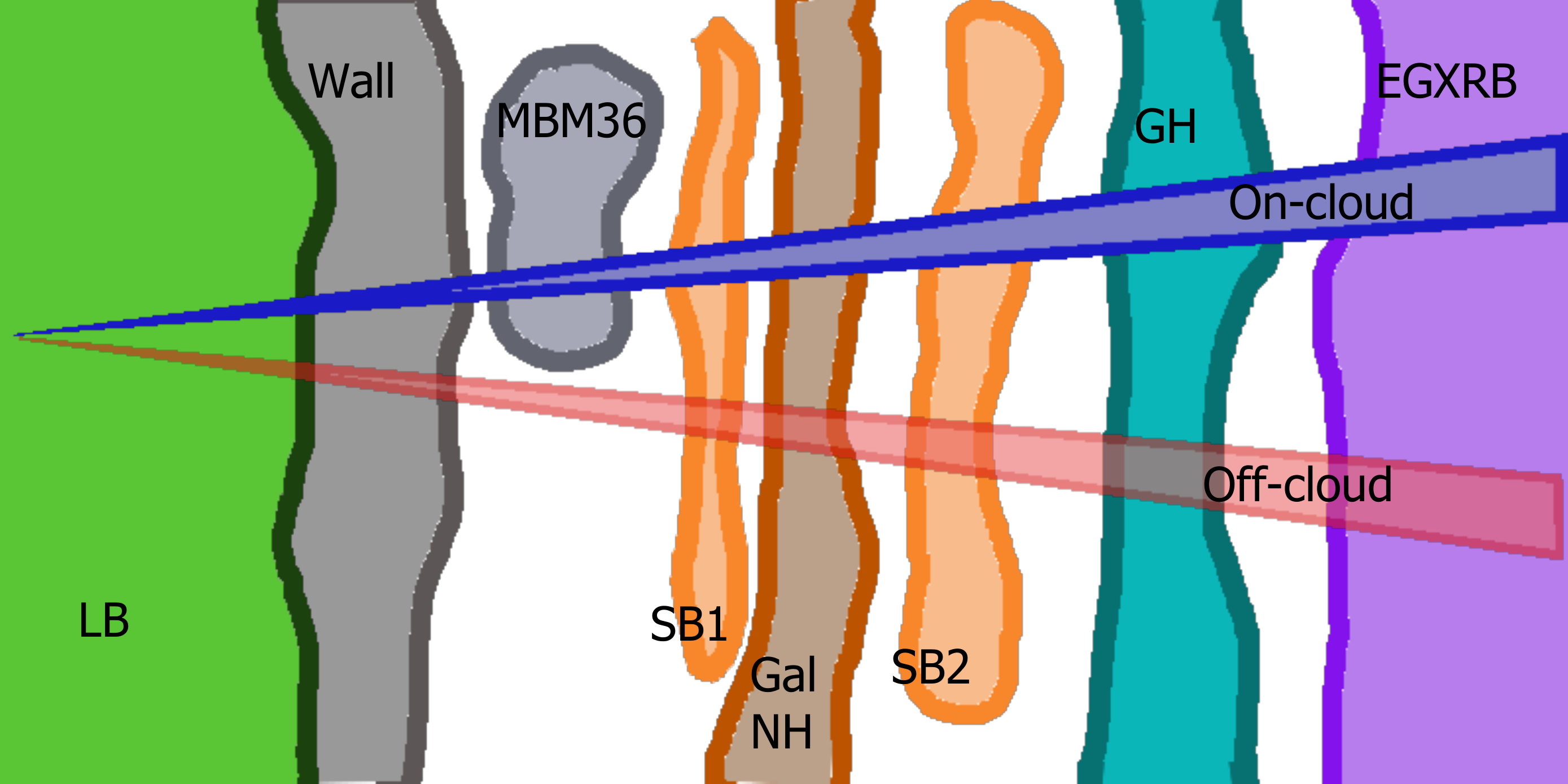}
\caption{Schematic representation of the contribution to the X-ray emission along the On- and Off-cloud targets line of sight. The LHB component includes the local contribution from within 100~pc (LHB and SWCX). A wall at $\sim100$~pc separates the LHB from the local ISM. MBM36 is located at $\sim140$~pc along the On-cloud line of sight. The SB predicted by local NPS models (SB1) should be right beyond MBM36 (strongly absorbed On-cloud, weakly absorbed Off-cloud). Beyond the ``wall'' due to galactic $N_H$ we find the SB predicted by non-local models (SB2) for NPS and the GH (distance larger than 1000~pc). The extragalactic X-ray backgrounds (EGXRB) originates from distances on the scale of several Mpc and more. Note: the size and distances of the contributors to the DXB and the angular separation between the lines of sight are not in scale.
\label{model_graph}}
\end{figure}

According to previous observations, there might be several components that contribute to the X-ray emission and absorption in the MBM36 region. Moving towards increasing distances (see the cartoon in figure~\ref{model_graph}), there is a local component (LHB plus SWCX) within $\sim100$~pc, an absorbing wall that separates the LHB from the near ISM, MBM36 at 140~pc (along the On-cloud line of sight), the SB predicted by local NPS models, a wall due to galactic $N_H$, the GH and the SB predicted by Galactic NPS models, and the extragalactic X-ray background. 

The local component includes contribution from both LHB and SWCX. Although recently two computational models for charge exchange have become available \citep{Cumbee14,Smith14}, the precise characterization of the local environment is beyond the scope of this work and therefore we used a simplified model. In the past, a single thermal equilibrium model \emph{apec} (\citealt{Foster12}, one of the most commonly adopted plasma model. In the remainder of this work we will name \emph{apec} the plasma model in general and not the \emph{XSPEC} code in particular) proved to be adequate to cumulatively describe the local components \citep{Galeazzi07, Henley08,Gupta09}, due to the poor energy resolution of the detectors. In order to improve this model and take in to account both the variations induced by SWCX in the O\elem{VII} bands and the O\elem{I} contamination described in section~\ref{Observation}, we model the local component using an \emph{apec} model with Oxygen abundance fixed at 0, and three $\delta$-functions at the energies of the O\elem{I} line (525~eV), of the O\elem{VII} K-$\alpha$ triplet centroid (567~eV), and of the O\elem{VII} K-$\beta$ line (666~eV). Notice that usually the O\elem{VIII} line at 654~eV is included in this modeling, and that it is almost indistinguishable from the O\elem{VII} K-$\beta$ line at CCD resolution. For this dataset, however, we obtained a better fit including only the O\elem{VII} K-$\beta$ line.

Based on the model represented in Fig.~\ref{model_graph} and on previous work in the same direction \citep{Willingale03, Miller08}, we expect at least two absorbed components. One component is associated with the GH/GB (as we will discuss later, mostly with the Galactic Bulge) and is absorbed by the neutral hydrogen column density, the other component is associated with the SB and in principle is only partially absorbed, depending on the actual location of the SB. Throughout this paper, we will refer to the fully absorbed component as GB since, as we will show, the GB component dominates the GH, and to the partially absorbed component as SB. In the case of SB1 (of figure~\ref{model_graph}) this components would be absorbed only by MBM36 and behave as a local component in the off-cloud pointing, while in the case of SB2, it would be fully absorbed. The requirement of a model with at least two absorbed thermal components was further corroborated by the poor results of a first attempt to fit a model with a single absorbed thermal component. 

We modeled the GB component with an absorbed thermal model.  The Off-cloud absorption is described by a single absorption model where the column density has been estimated using the \emph{IRAS100} values ($6.80\times10^{20}$~cm$^{-2}$). We used the absorption model by \citet{Morrison83}. On-cloud, instead, there are two sources of absorption, the ``galactic component'' and MBM36. In this case we adopt a combined absorption where the two absorption terms have column densities equivalent to the Off-cloud galactic value and to the MBM36 density. As previously noted (section~\ref{Observation}), the MBM36 column density is not well constrained and therefore we left it as a free parameter in the model. Indeed, we tested the GB model also using the on-cloud column density fixed at \emph{IRAS100}-estimated values but the fit was very poor.

The SB component is the most delicate to model since we have to carefully characterize the absorption factor. As shown in figure~\ref{model_graph}, the putative SB could be located either right outside the local wall or at several kpc away, near the Galactic Polar Axis. The associated absorption factor, therefore, could vary from a negligible value (very local SB) to a maximum equal to galactic absorption (if the SB is associated with an outflow from the Galactic Center). Accordingly, we do not use a  value fixed to the wall column density, but we leave it free to vary between 0 and full galactic absorption.

The extra-galactic background, finally, is modeled with an absorbed power law, where both exponential index and normalization are free parameters, and the absorption follows the same modeling as the GB.

This two-absorbed-thermal-components model is not yet adequate to fit the data, since the observed spectrum still shows an excess emission in the the band between 0.7~keV and 1.0~keV. This excess emission can be explained with an overabundance of Ne and/or Fe \citep{Yoshino09} or with the presence of an additional warmer thermal component. Both cases have been reported in literature. In order to account for the extra emission and discriminate between the two cases, therefore, we investigated two models, one with two absorbed thermal components with free metallicity for the most important metals, and one with three absorbed thermal components at fixed metallicity. We refer to the two models at 2T+ and 3T. The two models prove to have almost identical likelihood, however we favor the model with only two absorbed thermal components (therefore this is our fiducial model) for sake of simplicity, although the presence of a third component cannot be excluded. We summarize the two models in table~\ref{tab_models}). 

\begin{table}
\begin{center}
\caption{Summary of GH models. 
\label{tab_models}}
\begin{tabular}{|l|c|c|}
\hline
Component & Model 2T+ & Model 3T \\
\hline
SB & Fully absorbed &Fully absorbed \\
	& Free N & Fixed N \\
\hline
GB & Partially absorbed& Partially absorbed\\
\hline
T3 & - & Fully absorbed\\	    
\hline
\end{tabular}
\end{center}
\end{table}

We tested the two models performing combined fits of the On- and Off-cloud datasets. In addition, we limited the fits to the combined energy bands [$0.4-1.4$]~keV and [$2.3-7.0$]~keV to minimize the effects of non-Xray background.
 
\begin{figure*}
\plottwo{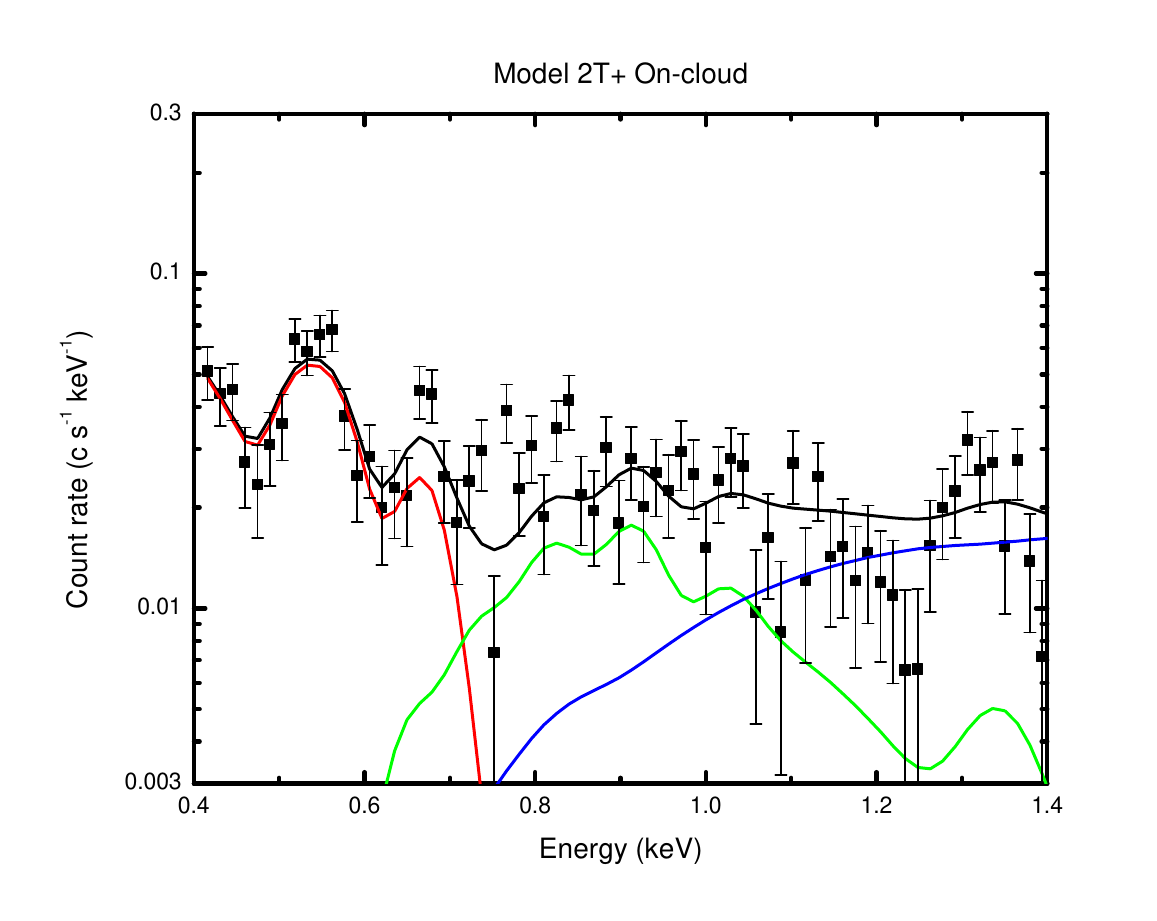}{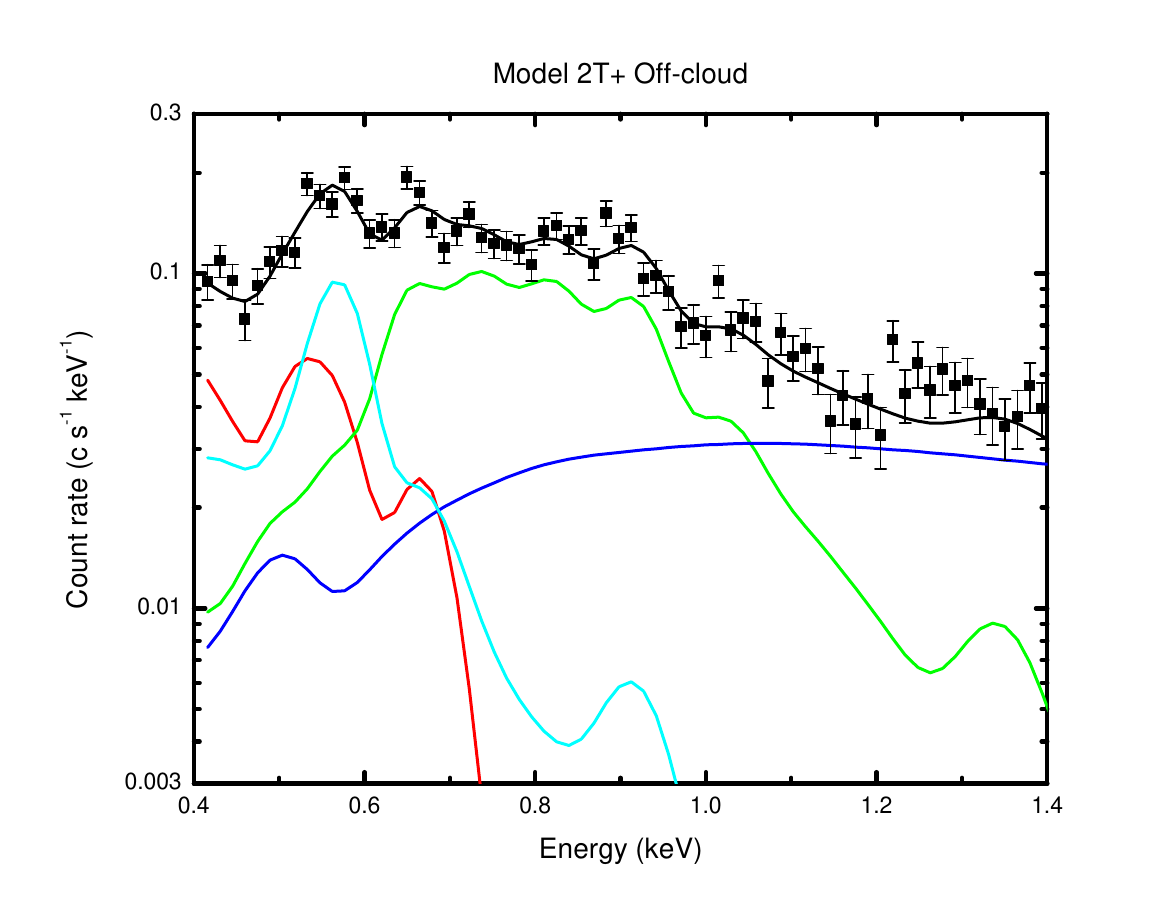}
\caption{MBM36-On (left panel) and Off (right panel) data (black squares) and folded models (black step line) for the the observation. The figure shows the spectrum  and the best fit  2T+ model folded through the response function, with the local (red line) component, the SB (green line), the GB (light blue line), and extragalactic sources (dark blue line).  
\label{plot_fits_01}}
\end{figure*}

In table~\ref{tab_fits} we report the best fit parameters for the two models.  We remind the reader that the column density of the Off-cloud ``MBM36'' absorption is set to 0, while it is left as a free parameter for the On-cloud target. Figure~\ref{plot_fits_01} shows the spectrum and the best fit of the 2T+ model.

\begin{table}
\begin{center}
\caption{Comparison table of the four GH models best fit parameters. 
\label{tab_fits}}
\begin{tabular}{|l|c|c|}
\hline Parameter & Model 2T+ & Model 3T \\
\hline
\hline
\rule{0pt}{3ex}  \raisebox{2pt}{$kT_{LHB}$ (keV)} & \raisebox{2pt}{$0.08^{+0.01}_{-0.08}$} &\raisebox{2pt}{$0.08^{+0.02}_{-0.08}$}\\ 
\rule{0pt}{3ex}  \raisebox{2pt}{$EM_{LHB}$ $(10^{-2}$~pc~cm$^{-6}$)} & \raisebox{2pt}{$6.4^{+0.8}_{-0.9}$} &\raisebox{2pt}{$6.1^{+1.0}_{-2.8}$}\\ 
\hline
\rule{0pt}{3ex}  \raisebox{2pt}{$Norm_{\elesm{O}{I}}$ (On) ($10^{-4}$)\tablenotemark{a}} &\raisebox{2pt}{$2.4^{+0.6}_{-0.6}$} &\raisebox{2pt}{$2.5^{+0.7}_{-0.7}$}\\ 
\rule{0pt}{3ex}  \raisebox{2pt}{$Norm_{\elesm{O}{I}}$ (Off) ($10^{-4}$)\tablenotemark{a}} &\raisebox{2pt}{$2.7^{+0.6}_{-0.6}$} &\raisebox{2pt}{$3.0^{+0.7}_{-0.7}$}\\ 
\rule{0pt}{3ex}  \raisebox{2pt}{$Norm_{\elesm{O}{VII K-$\alpha$}}$ ($10^{-4}$)\tablenotemark{a}} &\raisebox{2pt}{$4.4^{+1.0}_{-0.8}$} &\raisebox{2pt}{$4.4^{+1.0}_{-1.0}$}\\ 
\rule{0pt}{3ex}  \raisebox{2pt}{$Norm_{\elesm{O}{VII K-$\beta$}}$ ($10^{-4}$)\tablenotemark{a}} &\raisebox{2pt}{$1.5^{+0.3}_{-0.3}$} &\raisebox{2pt}{$1.5^{+0.4}_{-0.4}$}\\ 
\hline
\hline
\rule{0pt}{3ex}  \raisebox{2pt}{$N_{H\ MBM36}$ ($10^{20}$~cm$^{-2}$)} &\raisebox{2pt}{$51.1^{+3.4}_{-3.4}$} &\raisebox{2pt}{$51.1^{+3.6}_{-3.4}$}\\ 
\hline
\rule{0pt}{3ex}  \raisebox{2pt}{$N_{H\ wall}$ ($10^{20}$~cm$^{-2}$)} &\raisebox{2pt}{$<6.8$} &\raisebox{2pt}{$<6.8$}\\ 
\hline
\rule{0pt}{3ex}  \raisebox{2pt}{$kT_{SB}$ (keV)} &\raisebox{2pt}{$0.29^{+0.01}_{-0.02}$} &\raisebox{2pt}{$0.25^{+0.03}_{-0.03}$}\\ 
\rule{0pt}{3ex}  \raisebox{2pt}{$Ne_{SB}$} &\raisebox{2pt}{$1.7^{+0.3}_{-0.2}$}&\raisebox{2pt}{$1.0$}\\ 
\rule{0pt}{3ex}  \raisebox{2pt}{$EM_{SB}$ ($10^{-2}$~pc~cm$^{-6}$)} &\raisebox{2pt}{$1.0^{+0.1}_{-0.4}$} &\raisebox{2pt}{$1.1^{+0.1}_{-0.2}$}\\ 
\hline
\hline
\rule{0pt}{3ex}  \raisebox{2pt}{$\Gamma_{EGXRB}$} &\raisebox{2pt}{$1.62^{+0.05}_{-0.10}$}  &\raisebox{2pt}{$1.56^{+0.11}_{-0.12}$}\\ 
\rule{0pt}{3ex}  \raisebox{2pt}{$Norm_{EGXRB}$ ($10^{-3}$)\tablenotemark{b}} &\raisebox{2pt}{$1.06^{+0.13}_{-0.05}$}  &\raisebox{2pt}{$0.96^{+0.13}_{-0.12}$}\\ 
\hline
\hline
\rule{0pt}{3ex}  \raisebox{2pt}{$kT_{GB}$ (keV)} &\raisebox{2pt}{$0.11^{+0.01}_{-0.02}$}  &\raisebox{2pt}{$0.10^{+0.02}_{-0.03}$}\\ 
\rule{0pt}{3ex}  \raisebox{2pt}{$EM_{GB}$ ($10^{-2}$~pc~cm$^{-6}$)} &\raisebox{2pt}{$4.1^{+0.5}_{-1.4}$}  &\raisebox{2pt}{$6.2^{+1.5}_{-3.6}$}\\ 
\hline
\hline
\rule{0pt}{3ex}  \raisebox{2pt}{$kT_{T3}$ (keV)} &\raisebox{2pt}{-}  &\raisebox{2pt}{$0.76^{+0.14}_{-0.07}$}\\ 
\rule{0pt}{3ex}  \raisebox{2pt}{$EM_{T3}$ ($10^{-3}$~pc~cm$^{-6}$)} &\raisebox{2pt}{-}&\raisebox{2pt}{$1.7^{+0.6}_{-0.6}$}\\ 
\hline
\hline
\rule{0pt}{3ex}  \raisebox{2pt}{$\chi^2$/D.o.f.} &\raisebox{2pt}{$700/603$} &\raisebox{2pt}{$696/602$}\\ 
\hline
\end{tabular}
\tablenotetext{1}{Normalization assumes a solid angle with a 20' radius.}
\tablenotetext{2}{Surface brightness of the power-law model in the unit of photons~s$^{-1}$~cm$^{-2}$~keV$^{-1}$ at 1~keV for a 20' radius.}
\tablecomments{Error bars are at 90\% C.L.}
\end{center}
\end{table}

In both models the best fit temperature of the LHB is relatively low ($kT=0.081$~keV) compared to literature and the Emission Measure (EM) is $0.012$~cm$^{-5}$, comparable with \citet{Gupta09} but larger than standard references \citep{Snowden14}. These parameters are poorly constrained (the temperature is indeed stuck at the lowest limits of the \emph{apec} model) because we excluded the Oxygen lines from the analysis, and they are the best reference to estimate the temperature and density of the LHB plasma.

In our fiducial model, the GB component has a temperature similar to the LHB ($kT=0.11$~keV) while the SB component is significantly warmer ($kT=0.29$~kT). 

In order to account for the excess emission in the 2T+, we modeled the SB component leaving the most important metals (namely C, N, O, Mg, Ne, and Fe) as free parameters. The uncertainties on a global fit of all the metal abundances, however, are too large to provide reliable results and we had to constrain the abundances fitting one element at a time. In general Z/O (where Z is any metal) abundances ratios are all compatible with solar except for Ne, for which the ratio is Ne/O$\sim1.7$ times solar. The measured Ne excess does not depend on the O/H ratio, as we tested for different values of O/H (although we did not test metal distributions different from \citealt{Anders89}). For our fiducial model, therefore, we adopted fixed solar abundances for all metals, except for Ne that accounts for the excess emission. We also ruled out that the excess emission comes from the GB component since it is too cold and, in order to provide the photons required to fill the gap in the model, the metal abundance should be extremely high (Ne/O$>>10$ solar) and it would be hard to find a reasonable physical explanation. 

The best fit value of the MBM36 column density ($51.1\times10^{20}$~cm$^{-2}$) is roughly twice than the \emph{IRAS100} estimate, suggesting that the method describe in \citet{Snowden00} may not be reliable in the high column density regime. We also note that the $N_H$ value is about half of the $N_{H_2}$ (as can be inferred from figure~6 of \citealt{Juvela12}).

The ``wall'' absorption could not be constrained and we could only set an upper limit to the corresponding column density at the level of galactic absorption ($6.80\times10^{20}$~cm$^{-2}$). Without a precise estimate of the ``wall'' absorption we cannot use it to assess if the SB is only partially absorbed (and therefore located near the LHB) or if it is a large scale fully absorbed object.

\section{Discussion}
\label{discussion}

In our model the SB temperature is $kT=0.29$~keV, placing it in the upper range of the values reported by previous GH investigations and in good agreement with the reported ``NPS'' temperatures ($kT\approx0.3$~keV). The EM is 0.010~pc~cm$^{-6}$, well above the broad range reported for the GH by \citet{Henley13} so we believe it is, indeed, associated with excess emission associated with the NPS and not due to the GH. 

The GB has a temperature ($kT\approx0.11$) and the EM (0.019~pc~cm$^{-6}$) is 80\% larger than the SB one.  We discuss now the properties of the individual components  in the 2T+ model.

\subsection{The Galactic Bulge}
\label{discussion_gb}

The GB is the most prominent feature in the ROSAT 3/4~keV map together with the NPS. The X-ray emission from the GB has been extensively studied using both \emph{ROSAT} \citep{Snowden97, Almy00} and \emph{Suzaku} observations at low latitude \citep{Rocks08}. In particular, \citet{Rocks08} generated a very detailed model of the Bulge emission following the density and temperature profiles along the line of sight and splitting it in several smaller intervals. This approach is necessary for observations at very low latitude since both temperature and density have variations of about a factor 5 along the line of sight, thus greatly affecting the emission of the individual segments. At larger latitudes, instead, where the distance from the Galactic Center is more than 5~kpc (the line of sight toward MBM36, indeed, passes 6~kpc above the Galactic Center), the density and temperature variations are much smaller and we can approximate the line of sight as a single element with only one value of temperature and density all along the path. According to the model described by \citet{Almy00}, the Bulge contributes for about 100~R.U. at $l\sim35^\circ$. The model also predicts that at this latitude the Bulge temperature is $kT\approx0.13$~keV. Normalizing the emission for a plasma at $kT=0.13$~keV to the Bulge ROSAT contribution, we estimated $EM_{GB}=2.6\times10^{-2}$~pc~cm$^{-6}$. The spectrum of the GB contribution modeled on the \emph{ROSAT} data matches almost perfectly the GB component in our \emph{Suzaku} observation, making a very strong case that it is indeed the contribution from the Galactic Bulge.

\subsection{The SuperBubble}
\label{discussion_sb}
Since the SB emission is too strong to be simply associated with the ``generic'' GH, we investigated if a SuperBubble model could explain it. 

In order to model a local SB we can make use of the geometrical constraints described in \citet{Willingale03}. In this model the NPS is part of a ring centered at $l\approx352, b\approx15$ and radius $42^\circ$ and MBM36 is therefore located at $22^\circ$ away from the Loop center. By definition, the EM is the integral of the square of the density along the line of sight up to a distance $d$
\begin{equation}
EM=\int_0^d{n_e^2 dr},
\label{eq_em1}
\end{equation}
in the approximation that the density is constant, the distance $d$ can be computed as
\begin{equation}
d=\frac{EM}{n_e^2}.
\label{eq_em2}
\end{equation}
Assuming that the SB has uniform density equal to the LHB density ($n_e=4.7\times10^{-3}$~cm$^{-3}$ from \citealt{Snowden14}) and the EM computed with the fit, we estimate that the path through the SB along the MBM36 line of sight is $450^{+50}_{-180}$~pc. This path corresponds to a spherical bubble of radius $r\approx150$~pc, with the center located at a distance $d\approx280$~pc away from the Solar System (due to the linear relations involved in the computation, distances and radii have errors $+10\%, -25\%$). This model, however, is in contrast with at least two different sets of observations. First, \citet{Puspitarini14} provided a detailed map of the ISM within $\sim300$~pc from the Solar System and, towards the MBM36 direction, the ISM fills the space until 300~pc leaving no room for the SB described by our model (the closest edge would be $\sim130$~pc away). Besides, the SB has a thermal pressure about three times larger than the LHB. Assuming that the LHB and the SB are in similar environments and since the LHB is in pressure equilibrium, the pressure of the SB would be too high to be in equilibrium. The LHB thermal pressure ($\approx10600$~cm$^{-3}$~K) is larger than the interstellar thermal pressure ($\approx3200$~cm$^{-3}$~K) but is balanced by magnetic pressure of a $\sim4.9$~$\mu$G magnetic field \citep{Snowden14}. In order to balance the SB pressure, instead, the magnetic field should be $\sim10$~$\mu$G, much larger than the LHB environment and than the average values reported for the ISM, it is much closer to magnetic fields of dense clouds \citep{Ferriere01}, instead. 

If we keep the SB thermal pressure comparable to the LHB, with a density $n_e=1.6\times10^{-3}$~cm$^{-3}$, the corresponding path length inside the SB would be $3.9^{+0.4}_{-1.6}$~kpc, corresponding to a SB located $\sim3.8$~kpc away (half way to the Galactic Center) and of radius $r\sim2.4$~kpc. Such an object, however, has never been reported by other observations and cannot be explained using local scale models.

Considering the hypershell model, instead, the reduced density ($n_e=1.6\times10^{-3}$~cm$^{-3}$) is in good agreement with the prescription of \citet{Sofue00} that assumes a ``typical'' density of the order of $10^{-3}$~cm$^{-3}$. Assuming a very simplified model of a spherical shell, where the emission comes from both the layer between the Sun and the Galactic Center and the layer beyond the Galactic Center, the path length of 4~kpc correspond to a shell thickness of $\sim2$~kpc. This value matches the thickness estimated by \citet{Kataoka13}, based on the analysis of the changes of the emission measure crossing the upper side of the northern \emph{Fermi Bubble} (NPS region). The galactic scale suggested by this work, moreover, is also in agreement with the estimate of the distance of the NPS obtained using the absorption of molecular clouds in the Aquila Rift \citep{Sofue15}, set at a lower limit of $\sim1.0$~kpc. Although we have no further argument in favor of the association of the SB component with the hypershell model, our results show nice agreement with its predictions. 

\subsection{The Excess Emission}
\label{discussion_excess}

In the 2T+ model, the analysis of the SB metal abundances shows a moderately large fraction of Ne (Ne/O$=1.7$ solar). Ne overabundance, together with Fe overabundance, was also reported by \citet{Yoshino09}. In our analysis, however, we could not constrain the Fe abundance and we have been forced to set it at $\textrm{Fe/O}=1$ solar. Even if the Ne overabundance is statistically significant, it is nevertheless not too different from solar. There are several possibilities to explain the Ne overabundance. 

The first option is that the metallicity model adopted \citep{Anders89} is not adequate for the SB, other models can differ up to $\approx50\%$ in the ratio between O and other metals.

It has to be noted, however, that the angular resolution of \emph{Suzaku} is relatively poor and not all the point sources in the field can be removed. \citet{Kashyap92} showed that unresolved stars contribute to the X-ray background below 2.0~keV, in particular around $E=0.9$~keV. The Ne\elem{IX} excess, therefore, could be an artifact actually due to unresolved galactic stars. 

The third possibility is that Ne is the result of contamination from the Galactic Center activity along the line of sight. According to the ``hypershell'' model, starburst activity from the Galactic Center that occurred about $10^7$ years ago would be responsible of the large SuperBubble associated with the features of the NPS. The shock wave responsible of the heating of the gas within the SB should be rich in metals due to the high rate of star formation and supernovae explosions during starburst. While it works with Ne, this picture is not supported by the analysis of other metals, since the fitting procedure could not constrain their abundances and we had to force it at solar values. 

The last option is that Ne has standard solar abundance and that the extra emission instead comes from additional, warmer gas at $kT=0.76$~keV. In this case, however, the EM of the warm gas is very small and can be explained with the fact that we are observing a small object (clearly not the hypershell responsible of the NPS) or that the emission comes from a very thin layer of a much larger cloud (of the order of a few tens of parsecs). In the first case it becomes difficult to reconcile the source of this emission with sources responsible of the other Ne\elem{IX} detections. In the second case, instead, it is difficult to find a suitable model for a shell with both a very large inner radius and an extremely small thickness.

\section{Conclusions}
\label{conclusion}
In this work we have investigated the properties of the ISM in the direction of the high density molecular cloud MBM36. MBM36 is conveniently located inside the region defined by the NPS/Loop 1 structure and by the northern \emph{Fermi Bubble}. It can be used to assess if the emission of the NPS/Loop 1 region comes form a nearby SuperBubble, if it originates from a large hypershell above the Galactic Center, in the same scenario the describes the formation of the \emph{Fermi Bubbles}, or if it is just an isolated formation independent from any bubble/shell scenario. Thanks to the contrast between the cloud and a low density nearby target we have been able to separate the contribution to the DXB of local components (the Local Bubble and the SWCX) from that of distant source (both of Galactic and extragalactic origin). 

We identified two strong diffuse absorbed sources modeled with plasma at $kT=0.11$~keV and $0.29$~keV. Both thermal components have Emission Measure much larger than the typical values for the Galactic Halo emission. The colder thermal component is fully consistent with models of emission from the upper region of the Galactic Bulge. 

We ruled out the possibility that the emission from the warmer gas is due to a nearby SuperBubble, such a bubble would be located at less than 300~pc from the Solar System where independent observations \citep{Puspitarini14} have identified large dense regions, instead. Moreover, the SuperBubble would have a thermal pressure too high to be compensated by the thermal and magnetic pressures of the surrounding environment. This argument is also valid against models where the NPS structures corresponds to the emission from the shell of a nearby SuperBubble. The hypershell model, instead, is in good agreement with our observation. The temperature and emission measure of the warmer gas component, in fact, can be associated with a 2~kpc-thick shell generated by the shock front of starburst activity occurred near the Galactic Center $\sim10^7$ years ago.

We could not investigate the major GH models (Galactic Plane bound gas versus CircumGalactic Halo), the identified warmer component, in fact, is much brighter than any previously measured ``standard'' GH component.

We also identified excess emission corresponding to the energy band nearby the Ne\elem{IX} emission line (0.922~keV). The quality of the signal is not good enough to provide a definite interpretation of the excess. It is likely that the metallicity model is not adequate to describe the metal abundance in the Bulge region and that the excess Ne could arise from metal enrichment in the hypershell due to the starburst activity responsible of the shell itself. 

\acknowledgments

This work has been supported by NASA grants NNX11AF80G and NNX13AI04G. The authors would like to thank D. Koutroumpa, K. D. Kuntz, D. McCammon, and S. Snowden for the useful discussions and suggestions.

\end{document}